%Paper: astro-ph/9308036
%From: Yoshiaki Sofue <sofue@sof.mtk.ioa.s.u-tokyo.ac.jp>
%Date: Fri, 27 Aug 93 14:51:43 JST

%93 Aug; accepted for ApJ.Letter
%\magnification=1200
%\baselineskip=16pt

%%%%%%%%%%%%%%%%
%% Definitions:
%%%%%%%%%%%%%%%%
\def\sect{\vskip 2mm \centerline}

\def\r{\hangindent=1pc  \noindent}
\def\ref{\hangindent=1pc  \noindent}
\def\cen{\centerline}

\def\v{\vskip 1mm}

\def\noi{\noindent}
\def\kms{km s$^{-1}$}

\def\deg{$^\circ$}

\def\Vlsr{V_{\rm LSR}}

\def\tmb{$T_{\rm mb}$}
\def\Tmb{T_{\rm mb}}

\def\Ico{I_{\rm CO}}

\def\Msun{M_{\odot \hskip-5.2pt \bullet}}
\def\msun{$M_{\odot \hskip-5.2pt \bullet}$}

\def\halpha{H$\alpha$}

\def\Deg{^\circ}
\def\deg{$^\circ$}

\def\co{$^{12}$CO($J=1-0$)}

\def\htwo{H$_2$}

\def\Ta{T^*_{\rm A}}

\def\pa{PASJ}

\def\apj{ApJ}
\def\apjl{ApJ Letters}
\def\apjs{ApJS}
\def\aj{AJ}
\def\aa{A\&A}

\def\araa{ARAA}

\def\iaut{{\it Institute of Astronomy, University of Tokyo, Mitaka, Tokyo 181}}

\def\kiso{{\it Kiso Observatory, University of Tokyo, Kiso-gun, Nagano 397-01}}

\cen{\bf CO EMISSION FROM MOLECULAR CLOUDS IN}
\cen{\bf THE CENTRAL REGION OF M31}
\v
\cen{Yoshiaki SOFUE$^{1,2}$, and Shigeomi YOSHIDA$^1$}
\cen{1. \kiso, {\it Japan}}
\cen{2. \iaut, {\it Japan}}

\v\v
\cen{ABSTRACT}\v

We detected \co-line emission from a complex of dark cloud located at about 250
pc from M31's center, which is the most conspicuous extinction feature within
the central few hundred pc as observed in $B,V,R,I$-band CCD imaging.
The darkest cloud has a 30-pc size, and the virial mass is estimated to be
$\sim 8.7\times 10^5\Msun$, which is the most massive molecular cloud in the
bulge of M31.
The conversion factor from  CO intensity to \htwo column mass is found to be by
an order of magnitude greater than that for galactic molecular clouds.
The hydrogen mass-to-color excess ratio is also found to be anomalously large.

\v\noi SUBJECT HEADINGS: Dark clouds -- CO emission -- Galaxies; individual
(M31) -- Interstellar extinction -- Molecular clouds.

\sect{I. INTRODUCTION}\v

Allen and Lequeux (1993) detected CO-line emission from massive molecular
clouds of a few hundred pc size and a few $\times 10^7\Msun$ mass in the inner
disk of M31 at $\sim 2$ kpc distance from the nucleus.
Although there have been several attempts to map along the  major and minor
axes (Combes et al. 1977; Stark 1985;  Sandqvist et al. 1989), no detection of
CO has been reported in the central few hundred pc (e.g., Koper et al. 1991)
except for the earlier report of possible detection (Solomon and deZafra 1975).
HI observations also indicate a void in interstellar hydrogen (e.g., Brinks and
Shane 1984).
M31 exhibits a quantity of gas in its inner regions which is more
characteristic of early-type galaxies than of other Sb galaxies like the Milky
Way.
In fact, extensive surveys for optical dark clouds in the central region of M31
have revealed neither a disk nor spiral arms that are related to the outer disk
structure (Hodge 1980, 1981).

Recently, we performed a color-excess study of dark clouds in the central
region of M31 based on multi-color CCD data taken at the Kiso Observatory, and
reported detection of CO-line emission toward a complex of dark clouds in the
central region (Sofue et al. 1993).
In this paper, we discuss the physical properties of the clouds based on the
CO-line data from the Nobeyama 45-m telescope.

\sect{II. OBSERVATIONS}\v

We obtained CCD images filtered in the $B, ~V, ~R$, and $I$-bands on December
26, 1992  with the 1.05-m Schmidt telescope of the Kiso Observatory.
The frame with $1000\times 1018$-pixels covered a $12'.53\times 12'.76$ area.
The seeing size was about $4''$ (=13 pc).
We obtained several frames with different exposure times, and flat-fielding was
applied to each frame, which were finally added.
Intensity calibration was made by using star images in a nearby standard field.
The darkest area at the SE and NW corners 9$'$ from the nucleus was subtracted,
which yielded an error of about 5\% in the sense of underestimates of
intensity. This  error is found to yield an error in the color excess of about
$\sim 0.01$  mag..
 In Fig. 1 we  show a color-excess map obtained from $B$ and $V$-band frames:
the apparent color excess of a cloud is defined by
$ E_0(B-V)=(B-V)_{\rm cloud}-(B-V)_{\rm bg}, $
where subscripts `cloud' and `bg' denote values for the cloud and its
surrounding smooth background, respectively.

\cen{-- Fig. 1 --}

Conspicuous in Fig. 1 are dark clouds  at $\Delta {\rm RA} \sim -20''$ and
$\Delta {\rm Dec} \sim 75''$, where $\Delta {\rm RA}$ and $\Delta {\rm Dec}$
are offsets from the nucleus at RA = 00h 40m 00.1s, Dec = $40\Deg59'42'.7$
(1950)(NED 1992).
The outline of the clouds had been listed as D382, D384, and D395 by Hodge
(1981).
The size of each cloud is typically 10$''$ (33 pc in diameter).
The apparent color excess of the darkest cloud is measured to be
$ E_0(B-V)=0.10\pm 0.01$ and $ E_0(B-I)= 0.22\pm 0.01 $.
The intrinsic color excess due to a cloud, which can be related to interstellar
extinction, is derived by correcting for the foreground emission of the bulge,
which is assumed to contribute by a half to the total bulge emission.
Then, since $ E_0(B-V)$ and $ E_0(B-I) $ are small enough, the clouds' color
excess may be approximately double the apparent excess:
$ E(B-V)=0.20$ and $ E(B-I)= 0.44$.
Their ratio, $E(B-V)/E(B-I) \sim 0.45$, indicates a normal interstellar
extinction (Walker 1987).

A number of dark clouds and filaments are distributed in the central few
hundred pc (Fig. 1).
Their distribution traces a `` barred face-on'' spiral, apparently not related
to the outer disk [ see Sofue et al. (1993) for more detailed discussion of
optical features].
They also show a good coincidence with the off-plane face-on spirals observed
in the \halpha emission (Ciardullo et al. 1988).
These facts indicate that the dark clouds are off-plane objects, probably on a
plane perpendicular to the line of sight.
Then, the distance of the cloud from the nucleus is estimated to be 250 pc, and
its height from the galactic plane about 150 pc.
However, we cannot deny the possibility that the clouds lie in the disk plane,
in which case their distance from the nucleus is about 800 pc.

Deep \co-line observations of the central region were made using the Nobeyama
45-m telescope in December 1992 and February 1993 during the course of CO-line
study of nuclei in early-type galaxies.
The antenna had a HPBW of $15''$ (=50 pc) at 115.271 GHz, and the aperture and
 main-beam efficiencies were $\eta_{\rm a}=0.35$ and $\eta_{\rm mb}=0.50$,
respectively.
The intensity scale used in this paper is the main-beam brightness temperature
which is related to the antenna tempearture by
$\Tmb = \Ta / \eta_{\rm mb}$.
We used an SIS  receiver of system noise temperature of of about 500 K,
combined with a 2048-channel acousto-optical spectrometer with a velocity
coverage of 650 \kms.

We mapped six positions toward the dark-cloud complex as indicated by the
northern crosses in Fig. 1.
We also obtained several spectra along faint dark lanes near the nucleus at
PA=67\deg (central crosses in Fig. 1), which is associated with a nuclear mini
stellar bar (Sofue et al. 1993).
The reference off points were taken at $\pm5'$ east and west, where no
significant dark clouds was found.
The on-source total integration time was 1.5 hours per point, and baseline
fitting with third-order polynomials was applied.
After binding up every 32 channels, we obtained spectra with a velocity
resolution of 10 \kms and rms noise of $10-20$ mK \tmb.

\sect{III. RESULTS}\v

The obtained CO spectra toward the northern dark clouds are shown in Fig. 2a.
The CO line appears to have been detected nearly all positions at a velocity of
$\Vlsr \simeq -220$ \kms, although the detection is only 2 to 3$\sigma$ level.
In order to increase the signal-to-noise ratio, we integrated the six spectra
to get a composite spectrum as shown in Fig. 2b, which now shows the detection
at about 5$\sigma$ level.
The velocity agrees with that derived from the [OII] and [NeIII] lines at
approximately the same position (Ciardullo et al. 1988).
The velocity width at half maximum of the CO line is $\sigma_v \sim 30$ \kms,
and the integrated intensity is about $\Ico \sim 9\pm 2$ K \kms toward the
darkest cloud.
No significant emission was detected along the line crossing the nucleus (Fig.
2c).

\cen{-- Fig. 2 --}

The darkest cloud in the northern complex (Fig. 1) has an optical size of
$10''$ in diameter, or the radius is  $r \sim 17$ pc.
For the velocity width of $\sigma_v \sim 30$ \kms, the virial mass of the cloud
can be estimated as
$M_{\rm vir} \sim (\sigma_v/2)^2r /G \sim 8.7 \times 10^5 \Msun$,
where $G$ is the gravitational constant.
The total mass involved in the complex is estimated to be a few$\times
10^6\Msun$.
Thus, the dark cloud complex has a mass comparable to a giant molecular cloud.
As usual for a virial-mass estimation, the derived value contains an error of a
factor of two, which also applies to related values derived below.
We stress that the complex is most conspicuous, and therefore probably most
massive, within the central few hundred pc of M31's nucleus.
This shows a striking contrast to the nuclear disk of our Galaxy, where giant
molecular complexes by one or two orders of magnitude more massive have been
observed (e.g., Bally et al. 1987).

The column density of \htwo of this cloud is then estimated to be
$N_{\rm H_2} \sim M_{\rm vir}/(\pi r^2) \sim 6.3 \times 10^{22}$
\htwo cm$^{-2}$.
Since the beam area is slightly larger than the apparent optical size of the
cloud, the true CO intensity ($I_{\rm CO}$) can be estimated from observed
intensity  ($I_{\rm CO, obs}$) by correcting for the beam-dilution factor of
about $f \sim (10''/15'')^2$, and we obtain
$ I_{\rm CO} = I_{\rm CO, obs}/f \sim 20 $ K \kms.
{}From these, we derive a conversion factor from CO intensity to \htwo column
density as
$ X=N_{\rm H_2}/I_{\rm CO} \sim 3.1 \times 10^{21} $ \htwo cm$^{-2}$/K \kms (=
49 \msun pc$^{-2}$/K \kms).
This value is by a factor of ten greater than that  for molecular clouds in our
Galaxy as estimated from a similar virial-mass method:
$ X_{\rm Gal} \sim 3.6 \times 10^{20}$ \htwo cm$^{-2}$/K \kms (Sanders et al.
1984).

The gas-to-color excess ratio is obtained as
$  N_{\rm H_2} / E ( B - V ) \simeq  6.3 \times 10^{23}$ atoms cm$^{-2}{\rm
mag.}^{-1}$.
This is almost by two orders of magnitude greater than that for inter-cloud
value in the solar vicinity, $\sim 6 \times 10^{21}$ atoms cm$^{-2}{\rm
mag.}^{-1}$, which might increase significantly for dust in dense clouds
(Savage and Mathis 1979).

\sect{IV. CONCLUSION}\v

CO emission has been detected toward a dark cloud complex at 250 pc distance
from the center of M31, while no CO was detected toward the nucleus.
The darkest cloud in the complex has a size of about 30 pc and a virial mass of
$8.7 \times 10^5 \Msun$ (with an error of factor two), which is the most
massive cloud near the nucleus.
The extremely small amount of molecular gas in the central region shows a
contrast to the molecular-gas rich nuclear disk of our Galaxy.
The gas-to-CO intensity ratio, hence the conversion factor from CO
intensity-to-molecular hydrogen column mass, is estimated to be an order of
magnitude greater than the galactic value.
This indicates a smaller heavy element-to-gas ratio in the M31 center.
The gas-to-color excess ratio was found to be almost by two orders of magnitude
greater than the inter-cloud galactic value.
This might indicate that the dust-to-CO ratio in the bulge of M31 is
significantly smaller than that in the solar vicinity.

\sect{REFRENCES}\v

\r Allen, R. J., and Lequeux, J. 1993, \apjl, 410, L15.

\r Bally, J., Stark, A. A., Wilson, R. W., Henkel, C. 1987, \apjs,  65, 13.

\r Brinks, E. and Shane, W. W.  1984,  \aa, 55, 179.

\r Ciardullo, R., Rubin, V. C.,Jacoy, G. H., Ford, H. C., Ford, Jr., W. K.
1988, \aj, 95, 438.

\r Combes, F., Encrenaz, P., Lucas, R., and Weliachew, L. 1977, \aa, 61, L7.

\r Hodge, P. W. 1980, \aj, 85, 376.

\r Hodge, P. W. 1981, {\it Atlas of the Andormeda Galaxy} (University of
Washington Press, Seattle), chart 1.

\r Koper, E., Dame, T. M., Israel, F. P., Thaddeus, P. 1991, \apjl, 383, L11.

\r Sanders, D. B., Solomon, P. M., and Scoville, N. Z. 1984, \apj, 276, 182.

\r Sandqvist, Aa., Elfhang,T., Lindblad, P. O., 1989, \aa, 218, 39.

\r Savage, B. D. and Mathis, J. S. 1979, \araa, 17, 73.

\r Solomon, P. M., and deZafra, R. 1975, \apjl, 199, L79.

\r Sofue, Y.,  Yoshida, T., Aoki, T.,  Soyano, T., Tarusawa, K., and Wakamatsu,
K. 1993, \pa, submitted.

\r Stark, A. A. 1985, in {\it The Milky Way Galaxy}, ed. H. van Woerden et al.
(D. Reidel Publ. Co.), p. 445.

\r Walker, G. 1987, in {\it Astronomical Observations} (Cambridge University
Press), p. 17.

\vskip 30mm
\noi Figure Captions

\v\r Fig. 1: Color excess map of $E(B-V)$ for the central $4'\times 4'$
region. The contours are drawn at interval of 0.011 mag., starting at 0.011
mag., in excess over the background.
Note a dark-cloud complex at $1'.3$ (250 pc) north of the nucleus.
Crosses indicate positions of CO line observations, and the large cross
indicates the nucleus.

\v\r Fig. 2: (a) CO-line spectra observed toward the nuclear dark clouds for
positions indicated by the northern six crosses in Fig. 1b.
(b) Composite spectrum from (a).
(c) Spectra for positions near the nucleus indicated by the central six crosses
in Fig. 1.

\bye